\documentclass[osajnl,twocolumn,showpacs,superscriptaddress,10pt]{revtex4-1} 
\usepackage{amsmath,amssymb,graphicx}
\begin{document}

\title{Experimental realization of an optical antenna for collecting 99\% of photons from a quantum emitter}
%\textit{Optica }

\author{X.-L. Chu}
\affiliation{Max Planck Institute for the Science of Light, 91058 Erlangen, Germany}
\affiliation{Department of Physics, Friedrich Alexander University of Erlangen-N\"urnberg, 91058 Erlangen, Germany}

\author{T.J.K. Brenner}
\affiliation{Max Planck Institute for the Science of Light, 91058 Erlangen, Germany}
\affiliation{Institute of Physics \& Astronomy, University of Potsdam, 14476 Potsdam-Golm, Germany}

\author{X.-W. Chen}
\affiliation{Max Planck Institute for the Science of Light, 91058 Erlangen, Germany}
\affiliation{Department of Physics, Friedrich Alexander University of Erlangen-N\"urnberg, 91058 Erlangen, Germany}

\author{Y. Ghosh}
\affiliation{Materials Physics \& Applications: Center for Integrated Nanotechnologies, Los Alamos National Laboratory, Los Alamos, New Mexico 87545, USA}

\author{J. A. Hollingsworth}
\affiliation{Materials Physics \& Applications: Center for Integrated Nanotechnologies, Los Alamos National Laboratory, Los Alamos, New Mexico 87545, USA}

\author{V. Sandoghdar}
\affiliation{Max Planck Institute for the Science of Light, 91058 Erlangen, Germany}
\affiliation{Department of Physics, Friedrich Alexander University of Erlangen-N\"urnberg, 91058 Erlangen, Germany}

\author{S. G\"otzinger}
\affiliation{Department of Physics, Friedrich Alexander University of Erlangen-N\"urnberg, 91058 Erlangen, Germany}
\affiliation{Max Planck Institute for the Science of Light, 91058 Erlangen, Germany}
%\author{}\email{Corresponding author: xyx@osa.org}

\begin{abstract} We present the fabrication and characterization of an optical antenna that converts the dipolar radiation of a quantum emitter to a directional beam with 99\% efficiency. Aside from its implications for efficient detection of nanoscopic emitters, this antenna facilitates a deterministic single-photon source with applications in quantum information processing, metrology and sub-shot-noise detection of absorption. We discuss the photophysical limitations of the currently used quantum emitters for the realization of such a device.
\end{abstract}

%\ocis{ }

\maketitle

There is a great deal of interest in the optics community for efficient collection of light from atoms, molecules, quantum dots or other nanoscopic emitters \cite{obrien07,Scheel:09,Buckley:12}. One of the motivations concerns the detection of weak traces of fluorescence in biophotonics, where studies of chromophores with low quantum yield \cite{Lakowicz:04} or imaging of fast dynamical phenomena \cite{Kukura:09b,Hsieh:14} are desirable. Similarly, spectroscopy of weakly-emitting quantum systems such as rare earth ions \cite{Utikal:14} and color centers can benefit from higher count rates. Another fascinating prospect is in the realization of a new primary intensity standard based on a known flux of single photons obtained from an isolated quantum emitter through pulsed excitation \cite{Cheung:07}. A very different area of interest is in quantum optics with prominent applications in sub-shot-noise detection \cite{lounis05}, quantum communication \cite{Gisin:02}, and quantum computation \cite{obrien07}. In these applications losses quickly compromise the performance. For example, it is often stated that efficiencies on the order of 99\% are required for practical quantum computing \cite{Buckley:12}. In quantum key distribution, low losses are desirable both for increased data transmission rates and for higher security of the connection. 

Many different approaches have been explored for achieving high collection efficiency from single emitters. For example, a deep parabolic mirror has been used to extract light from an atom in a trap \cite{Golla:12}, however this solution is incompatible with most applications in the condensed phase. Another strategy has been to manipulate the density of states in microcavities \cite{barnes02}, but the need for operation under resonance conditions and the difficulty of fabrication have hampered collection efficiencies close to unity. Novel geometries using nanowires and their combination with cavity arrangements also hold promise for high collection efficiency \cite{claudon10,khajavikhan_thresholdless_2012}, but practical combination of these solutions with generic emitters is nontrivial and yet to be demonstrated. Recently, plasmonic nanostructures have also been combined with concepts from antenna theory and near-field optics to introduce directionality to the radiation pattern of emitters \cite{kuehn08, curto10}. Unfortunately, however, near-field coupling to metals is intrinsically accompanied by substantial losses. 

Three years ago, we devised a new antenna concept based on planar dielectric structures \cite{lee11planar}. In that work, we experimentally demonstrated 96\% collection efficiency from a single molecule that was oriented normal to the antenna plane. The noise on the light generated by a perfect emitter in such an antenna is reduced to 20\% of the shot noise value. This can be deduced from $\sqrt{xN(1-x)}$, which describes the fluctuation in the number of detected photons with an average value $N$ and loss factor $x$ \cite{lounis05}. Improvements beyond a certain noise level become increasingly more difficult to achieve. For example, a collection efficiency of 99\% would reduce the noise to only about 10\% of the shot noise value, corresponding to 10 dB intensity squeezing.

Motivated by the need for low-noise light sources and the desire to generalize our planar dielectric antenna to arbitrarily oriented emitters, we proposed a metallo-dielectric antenna design for reaching collection efficiencies exceeding 99\% for any orientation of the emitter dipole moment \cite{Chen:11}. In this article, we report on the first fabrication and characterization of such an antenna and elaborate on the subtleties involved in the realization of a single-photon source with 99\% efficiency. 

\begin{figure}[t]
\includegraphics[width = 0.99\columnwidth]{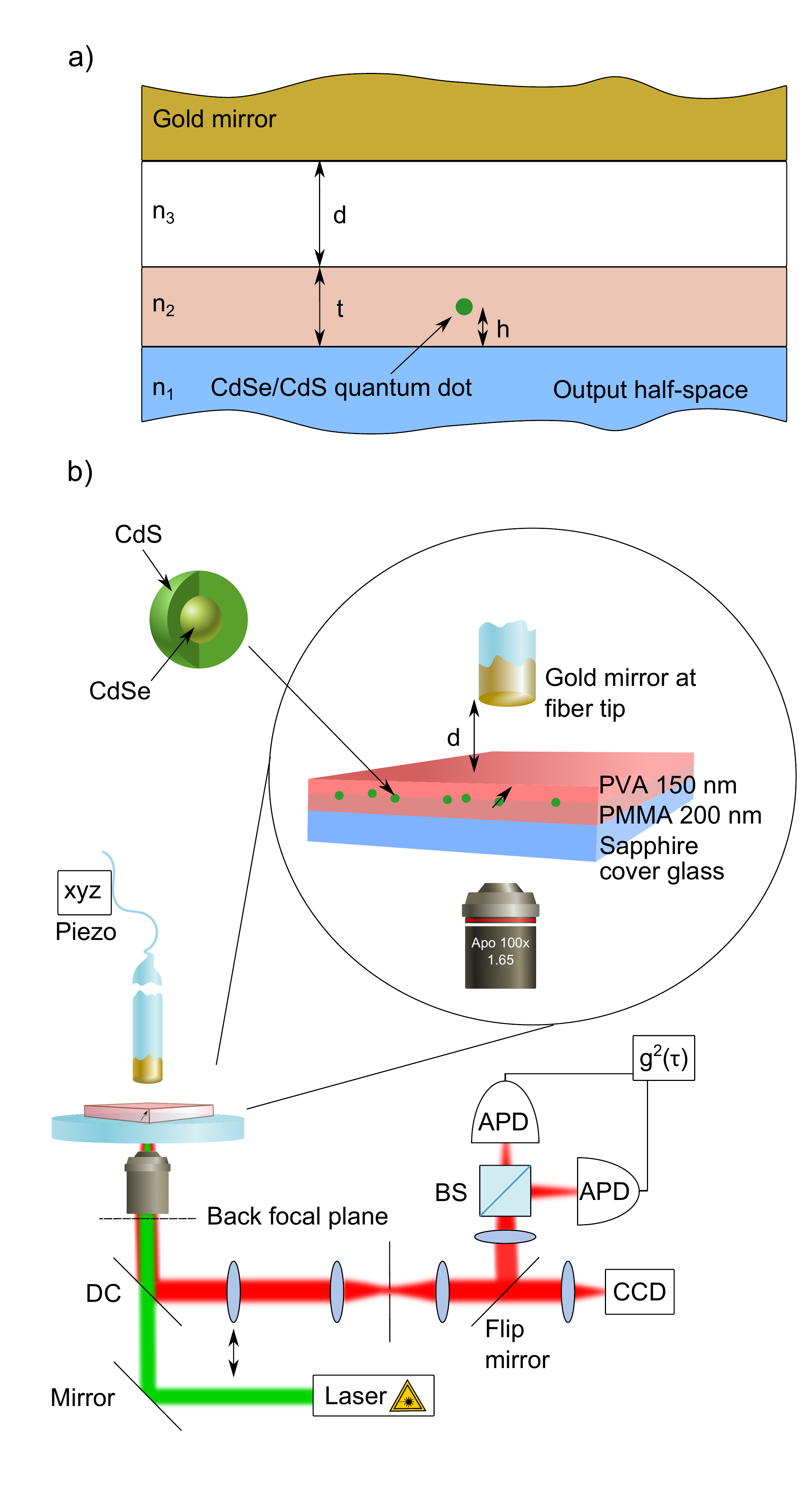}
\caption{a) The layer arrangement of the metallo-dielectric antenna.  b) Schematic diagram of the experimental setup. CdSe/CdS giant quantum dots are embedded in between two polymer layers (PMMA and PVA) of the same index of refraction, and a gold-coated end of a fiber is placed at variable separations from the top. A laser beam (wavelength 532 nm) illuminates the sample through a 100x microscope objective. The emitted fluorescence is collected by the same objective and guided to two avalanche photodiodes to measure the second-order correlation function $g^2(\tau)$ or to a CCD camera for back-focal-plane imaging. BS: beam splitter, DC: dichromatic mirror. }
\label{schematics}
\end{figure}

Figure \ref{schematics}a sketches the main ingredients of a planar metallo-dielectric antenna. The general requirement of $n_1>n_2>n_3$ remains the same as the previously-discussed dielectric antenna \cite{lee11planar}. However, to capture and redirect any leakage in the direction normal to the antenna plane, we add a mirror on the far side of the $n_3$ medium. The simplest experimental realization of this structure is to use a metallic mirror at a large enough distance from the $n_2-n_3$ interface to avoid coupling to surface plasmons. 

In this work, we have chosen to perform our experiments with semiconductor nanocrystals. In particular, we have used ``giant" CdSe/CdS core/thick-shell quantum dots \cite{Chen:08} featuring nearly complete suppression of blinking and fluorescence intermittency \cite{Ghosh:12}. Figure \ref{schematics}b displays the schematics of our experimental arrangement. The first component of the antenna consists of a sapphire cover glass (refractive index $n_1=1.78$).  We then coat this substrate with 200 nm of PMMA (refractive index 1.49), and spin cast quantum dots on this layer at a very small surface coverage. Next, a 150 nm thick layer of PVA (refractive index 1.5) is coated on top. Considering the similarity of the refractive indecies of PVA and PMMA, we can treat these combined layers as the second antenna medium with $n_2=1.5$. The thickness of this medium is measured using atomic force microscopy. The third medium consists of air with $n_3=1$.  In this proof-of-principle demonstration, we have decided to place the metallic mirror at the cleaved end of a thinned optical fiber so that we could vary its position at will. Quantum dots are excited by a laser at $\lambda=532$ nm (CW or pulsed), and their fluorescence is collected using the same microscope objective (Olympus Apo 100x, NA 1.65). The resulting emission centered at the wavelength of 637 nm is detected on a CCD camera or analyzed using a Hanbury-Brown and Twiss coincidence setup.

\begin{figure}
\includegraphics[ width  = 0.99\columnwidth]{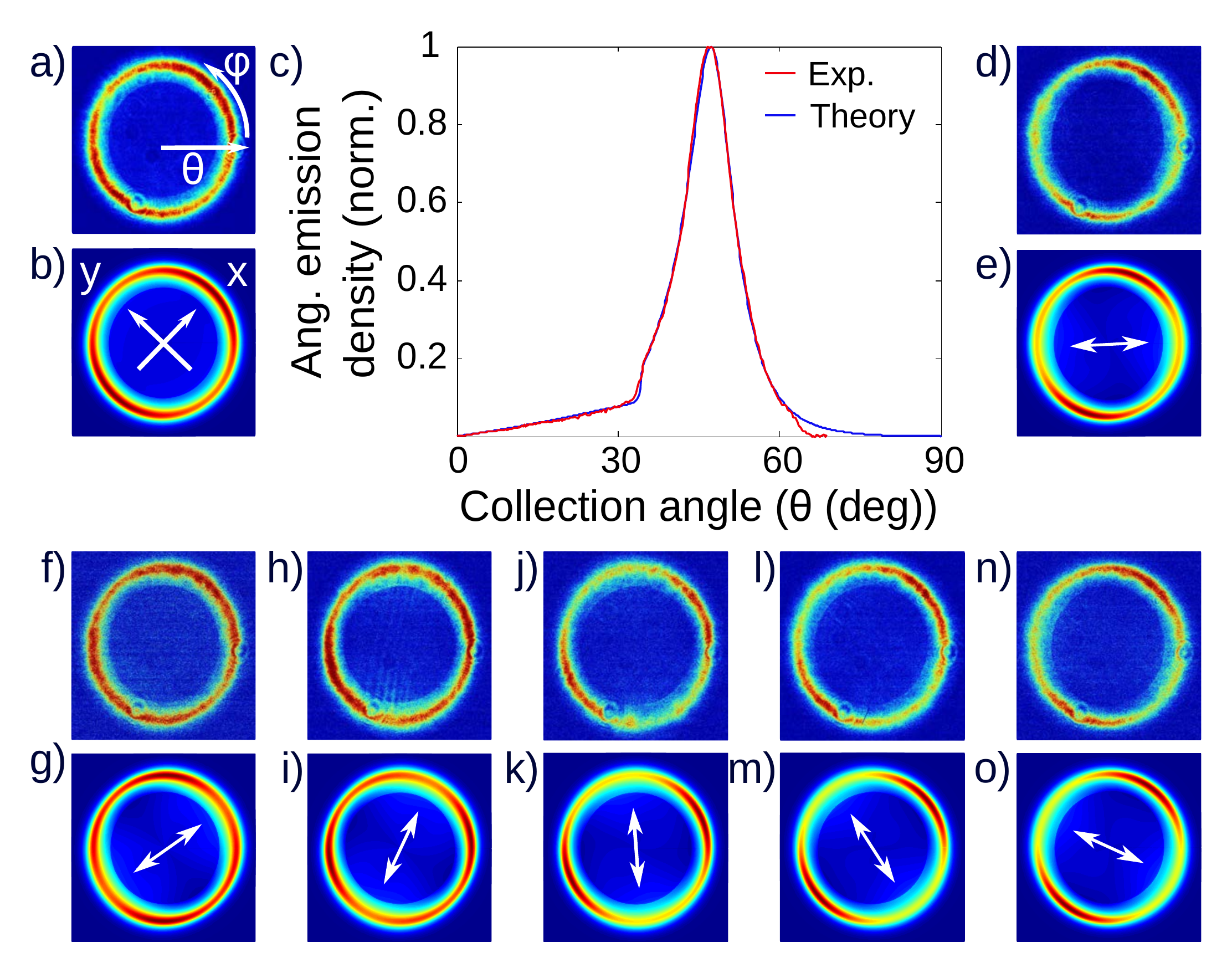}
\caption{Characterization of the angular emission from a single quantum dot in a dielectric antenna before the gold mirror was added. a) Power distribution of the emission in the back focal plane of the microscope objective. b) Theoretical fit to the image in (a), yielding information about the emission dipole moments (see text for more information). c) The signal in (a) integrated over the angle $\phi$ plotted as a function of the polar angle $\theta$ as indicated in (a). d, f, h, j, l, n) Experimental measurements as in (a) but for six different settings of a polarizer angle in the detection path at $30^\circ$ increments. e, g, i, k, m, o) The corresponding theoretical predictions (not fits). The white arrows indicate the orientation of the polarizer in each case.}
\label{bfp_pol}
\end{figure}

The key function of the antenna is to redistribute the radiated power by an emitter into a solid angle that can be collected by a commercial microscope objective. To investigate the angular emission pattern of a single quantum dot, we imaged its intensity distribution at the back focal plane of the objective. Figures \ref{bfp_pol}a and \ref{bfp_pol}b display the experimental outcome and its theoretical fit, respectively, for the dielectric part of the antenna without the gold mirror. 
\begin{figure*}[htbp]
\includegraphics[ width  = 0.99\textwidth]{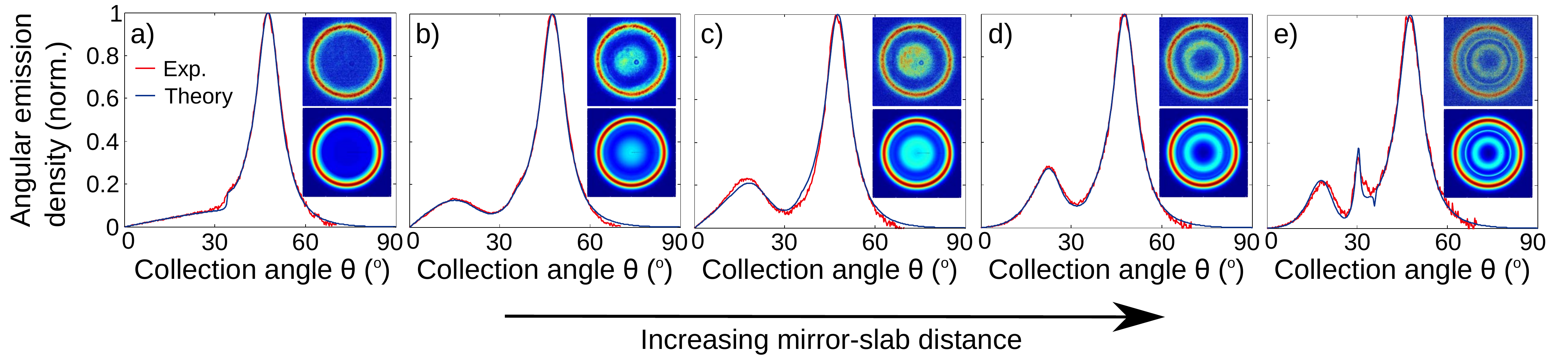}
\caption{Theoretical and measured angular fluorescence signal of a single quantum dot in the antenna structure. Insets: measured (top) back focal plane image and theoretical (bottom) simulations.  a) no metallic mirror. b-e) in the presence of a gold mirror placed at distances of $d=225$ nm, 284 nm, 355, and 680 nm, respectively, from the top surface of the dielectric antenna.}
\label{backfocal}
\end{figure*}
To arrive at Fig. \ref{bfp_pol}b, we first determined the ratio between the axial and in-plane dipole contributions by fitting the data in Fig. \ref{bfp_pol}c, which plots the detected power integrated over the azimuthal angle $\phi$ as a function of the polar emission angle $\theta$ (see Fig. \ref{bfp_pol}a). The agreement between the experimental (red) and theoretical (blue) traces is excellent. The slight deviation at angles beyond $62^\circ$ is caused by the unfortunate position of a small scattering impurity on the detection optics (see for example Fig. \ref{bfp_pol}f.). Once the in-plane dipole contribution was extracted, it was further separated into x and y components by fitting the image in  Fig. \ref{bfp_pol}a. We remark that semiconductor nanocrystals are known to allow for more than one independent dipole moment \cite{Chung:03}. In the case presented in Fig. \ref{bfp_pol}, the axial, x, and y components amounted to 44\%, 21\% and 35\%, respectively.  

Figures \ref{bfp_pol}d-o present the polarization dependence of the emission pattern obtained by placing a polarizer in the detection path. In Figs. \ref{bfp_pol}d, f, h, j, l, n we incremented the orientation of the polarizer in steps of $30^\circ$. Figures \ref{bfp_pol}e, g, i, k, m, o plot the corresponding theoretical predictions based on the dipole orientations deduced from the fit in Fig. \ref{bfp_pol}b. The excellent agreement between the measured and calculated angular intensity distributions for all polarization projections without additional fitting confirms the robustness of our analysis and the assignment of the quantum dot emission dipoles.

The insets in Fig. \ref{backfocal}a display the experimental (top) back focal plane images of another quantum dot without the gold mirror (i.e. similar to the case in Fig. \ref{bfp_pol}a) as well as the theoretical fit to it (bottom), while the main graph in the figure shows the integration over $\phi$. We now introduce the gold-coated end of an optical fiber with a diameter of about 15 $\mu$m. By using a piezoelectric scanner, we could place the mirror at different separations from the dielectric part of the antenna. The top insets in Figs. \ref{backfocal}b-e display the measured back focal plane fluorescence distribution for separations of 225 nm, 284 nm, 355 nm and 680 nm. The lower insets present the theoretical predictions. We emphasize that once the data in Fig.  \ref{backfocal}a were fitted, the following calculated images were obtained by merely considering a gold mirror at the corresponding measured separation. The main parts of the figures show the detected power integrated over $\phi$ as a function of $\theta$. These data reveal that for large mirror-slab distance, the part of the quantum dot emission reflected from the mirror interferes with the downward emission to yield the observed modulations. The remarkably good agreement between the experiment (red) and theory (blue) verifies the predicted contribution of the metallic mirror. 

To quantify the effect of the gold mirror further, we compared the photon flux with and without it. The leakage out of the far side of the dielectric part of the antenna amounts to 4\% for an axially-oriented dipole and 12\% for dipole moments in the antenna plane. We found that the detected fluorescence signal was consistently 10\% larger in the presence of the mirror. This is in very good agreement with the expected effect of 9.5\% for this quantum dot, which had 31\%  and 69\% of its emission dipole in the axial and planar components, respectively. We note that the quantum dot was excited via total internal reflection through the microscope objective to avoid fluorescence from the gold mirror. 

The studies presented above let us conclude that the antenna fully meets its design specification, chosen to yield more than 99\% collection efficiency within a half angle of $68^\circ$ accessible to a microscope objective with a numerical aperture of 1.65. This addresses an outstanding bottleneck in the realization of deterministic single-photon sources. What is now required is to excite a quantum emitter using short pulses at a well-defined repetition rate. However, a successful executation of this scheme also requires an emitter that has a fluorescence stability better than 99\%. An ideal two-level atom would satisfy this condition because it emits as many photons per unit time as excitation pulses if these were spaced sufficiently far apart with respect to the spontaneous emission lifetime. Unfortunately, a thorough examination of the existing quantum emitters lead us to the conclusion that this task is currently beyond reach due to limited photophysical properties. 

Over the past two decades, molecules, quantum dots and color centers have been explored as single-photon sources \cite{lounis05}, whereby the term single-photon source has become unanimous with a system that never emits two photons at a time. However, scientists are yet to demonstrate a source that would deliver a photon at an expected time. Several issues have to be considered, and in each system one or more of these pose problems. First, a non-unity quantum yield would introduce randomly missing photons. This issue is exacerbated by the difficulty of determining quantum efficiencies with one percent accuracy. Second, limited photostability is prohibitive for practical applications and long-term usage. Third, at high excitation intensities the supporting medium often fluoresces at the level of a few percent. Fourth, intermittent transitions such as intersystem crossing or blinking introduce randomness. Fifth, multiphoton absorption or multiexcitonic behavior might compromise the two-level character of the emitter at high excitation rates. 

Conventional dye molecules fall short of these requirements because of bleaching and blinking although selected systems have been shown to be unusually photostable \cite{pfab:04,Toninelli:10}. Atoms and ions in the gas phase are not favorable because of the difficulty of trapping and integration into compact devices. Color centers in bulk diamond can be extremely photostable \cite{Jelezko:06}, but they also suffer from intersystem crossing and blinking. In addition, color centers in nanocrystals have a large inhomogeneity in terms of quantum yield and fluorescence lifetime. Another alternative has been discussed in the context of semiconductor quantum dots. Here one usually distinguishes between epitaxially-grown nanostructures and chemically produced nanocrystals such as CdS/CdSe core-shell dots. The former offer excellent suppression of two-photon emission, but they might confront various undesired states involving dark or charged excitons \cite{Michler:03,Strauf:07}. Furthermore, good optical properties of these dots only emerge at cryogenic temperatures. Standard core-shell nanocrystals suffer from severe blinking, which makes precise power calibration impossible. Giant core-shell nanocrystals solve the blinking problem by suppressing Auger recombination, but at high pump intensities they can undergo transitions to biexciton and higher-order multiexciton states, resulting in deviations from a strong two-photon emission suppression.  

\begin{figure}[h]
\includegraphics[ width  = 0.99\columnwidth]{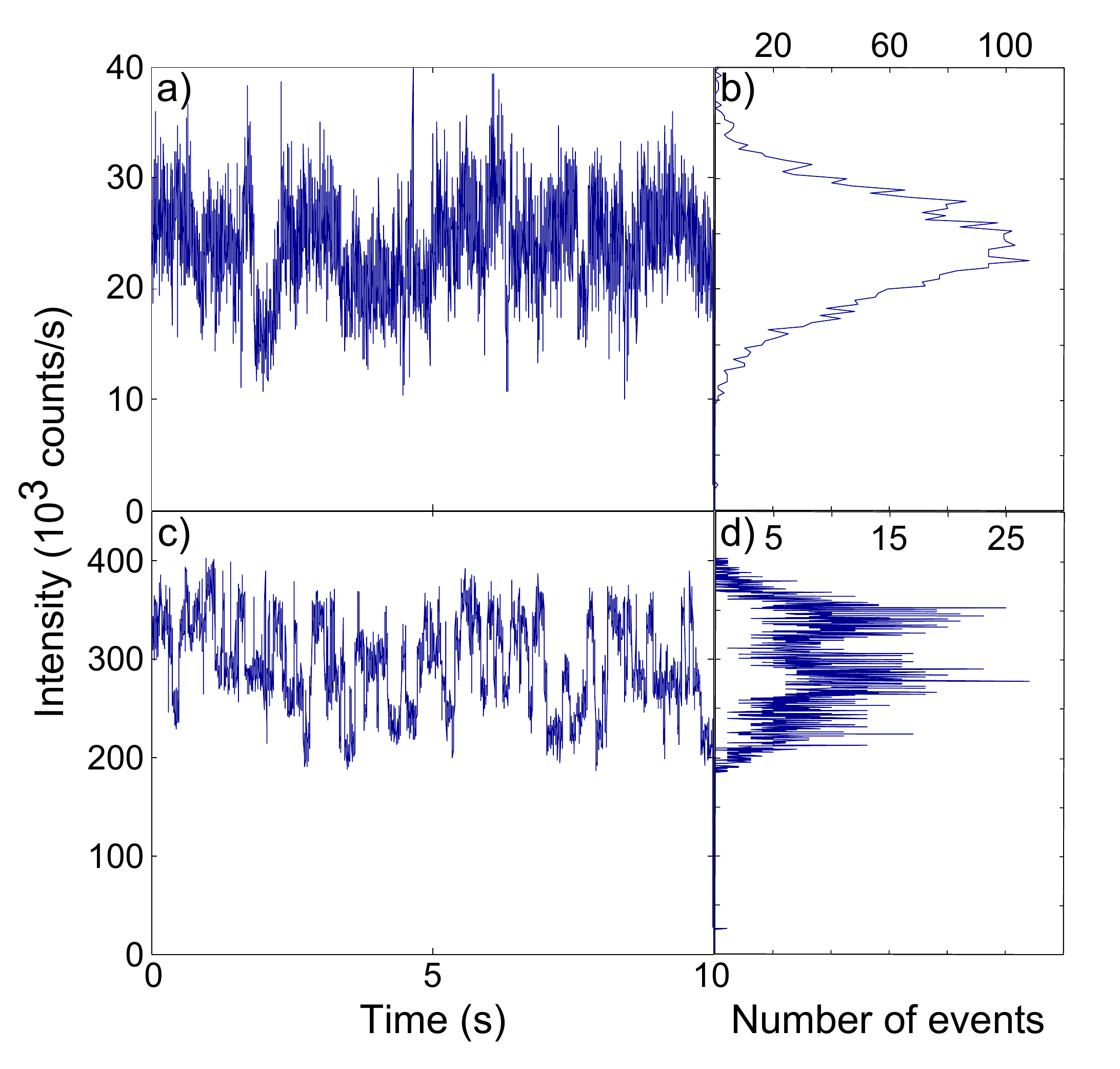}
\caption{a, b) Time traces of a single quantum dot with 1 ms resolution and 10 s acquisition time under an excitation power of 40 nW (a) and 1$\mu$W (b). c, d) Distribution of the fluorescence signal corresponding to (a) and (b), respectively.}
\label{timetrace}
\end{figure}

\begin{figure}
\includegraphics[ width  = 0.99\columnwidth]{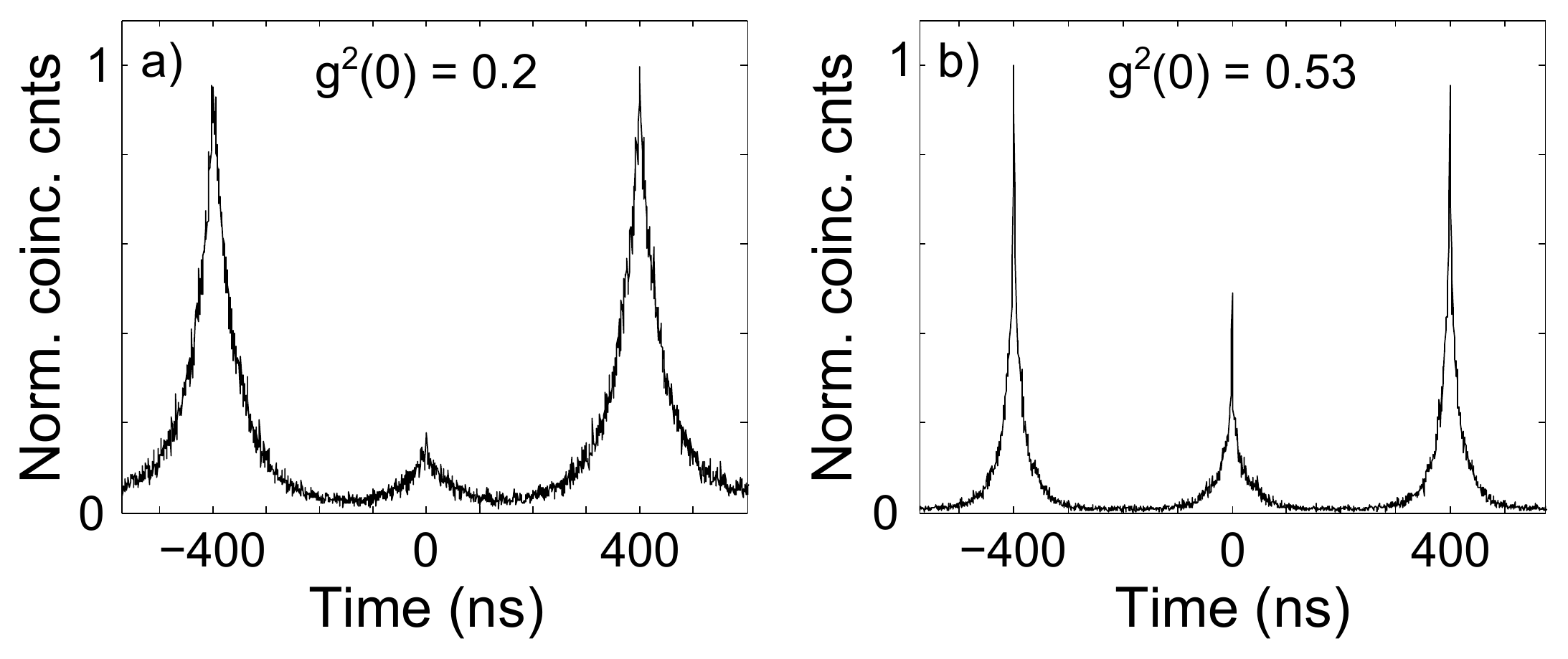}
\caption{Antibunching curves recorded from a quantum dot coupled to the antenna at excitation (pulsed) powers of 40 nW(a) and 1.19$\mu$W (b). The bin size was 3.2 ns.}
\label{satcurves}
\end{figure}

In the current work, we have chosen the giant quantum dots for their strong photostability and access to several dipole moment components. Although these dots do not blink in the common sense of experiencing on-off states, Figures \ref{timetrace}a and b show that the fluorescence can fluctuate about a mean value at weak excitation and flicker among different on-levels \cite{Galland:11} under strong illumination. Furthermore, it has been previously shown that significant suppression of Auger recombination in giant quantum dots leads to efficient room-temperature biexciton \cite{Park:11} and multiple-exciton emission \cite{Htoon:10}. Figure \ref{satcurves} displays the second-order autocorrelation function determined from Hanbury-Brown and Twiss measurements at low (a) and high (b) excitation powers. While antibunching is clearly observed in the former case, $g^{(2)}(\tau=0)$ grows above 0.5 at high powers. We note that we have verified that the reduction of the antibunching is not due to residual background. 

In conclusion, we have demonstrated a planar metallo-dielectric antenna that redistributes the photons from an emitter in excellent agreement with its theoretical design, which predicts a collection efficiency larger than 99\%. Such an antenna provides a crucial building block of an ultrabright single-photon source that can deliver up to several tens of million photons per second at deterministic times. Here, it will be crucial to reduce losses in the emission and detection processes also to the per cent level. Athough the photophysics of quantum emitters and the quantum yield of detectors currently do not meet this standard, the enormous recent progress in the development of efficient single-photon counters \cite{Marsili:13} and the synthesis of novel emitters \cite{Mangum:14, Mangum:14b} promise to address these issues. 

\smallskip

%\section*{Funding Information}
This work was funded by a European Research Council Advanced Grant (SINGLEION), the Max Planck Society and by project SIQUTE (contract EXL02) of the European Metrology Research Programme (EMRP). Giant quantum dots were synthesized at the Center for Integrated Nanotechnologies, a U.S. Department of Energy, Office of Basic Energy Sciences user facility. Los Alamos National Laboratory, an affirmative action equal opportunity employer, is operated by Los Alamos National Security, LLC, for the National Nuclear Security Administration of the U.S. Department of Energy under contract DE-AC52-06NA25396.

%\bibliography{vahid}

%

\end{document}